\documentclass[prd,twocolumn,superscriptaddress,showpacs,amsmath,amssymb]{revtex4}
\usepackage[below]{placeins}
\usepackage{amsmath}
\usepackage{graphicx}
\begin{document}
\title{\textbf{Bulk viscous cosmology: statefinder and entropy}}                                
\author{Ming-Guang Hu$^2$
\footnote{Email: huphys@hotmail.com}
and
 Xin-He Meng} \altaffiliation{Communication author}
\email{xhm@nankai.edu.cn} \affiliation{CCAST (World Lab), P.O.Box
8730, Beijing 100080, China} \affiliation{Department of physics,
Nankai University, Tianjin 300071, China (post address)}
\date{\today}
\begin{abstract}
The statefinder diagnostic pair is adopted to differentiate
viscous cosmology models and it is found that the trajectories of
these viscous cosmology models on the statefinder pair $s-r$ plane
are quite different from those of the corresponding non-viscous
cases. Particularly for the quiessence model, the singular
properties of state parameter $w=-1$ are obviously demonstrated on
the statefinder diagnostic pair planes. We then discuss the
entropy of the viscous / dissipative cosmology system which may be
more practical to describe the present cosmic observations as the
perfect fluid is just a global approximation to the complicated
cosmic media in current universe evolution. When the bulk
viscosity takes the form of $\zeta=\zeta_{1}\dot{a}/a$($\zeta_{1}$
is constant), the relationship between the entropy $S$ and the
redshift $z$ is explicitly given out. We find that the entropy of
the viscous cosmology is always increasing and consistent with the
thermodynamics arrow of time for the universe evolution. With the
parameter constraints from fitting to the 157 gold data of
supernova observations, it is demonstrated that this viscous
cosmology model is rather well consistent to the observational
data at the lower redshifts, and together with the diagnostic
statefinder pair analysis it is concluded that the viscous cosmic
models tend to the favored $\Lambda$CDM model in the later cosmic
evolution, agreeable to lots of cosmological simulation results,
especially to the fact of confidently observed current
accelerating cosmic expansion.
\end{abstract}
 \maketitle
\section{Introduction}
Observations of type Ia supernova(SNe Ia) suggest that the
expansion of the universe at later stage  is in an accelerating
phase. Additionally, the measurement of the cosmic microwave
background (CMB) \cite{DNS} and the galaxy power spectrum
\cite{MT} indicate that in spatially flat isotropic universe,
about two-thirds of the critical energy density seems to be stored
in a dark energy component (the simplest candidate is the famous cosmological
constant $\Lambda$) with negative enough pressure
\cite{AGR0}. Ironically, we do not know much about dark energy (DE)
properties, if not less than those on the mysterious dark side of
the universe \cite{sc}.

In order to explore the implying accelerating mechanism, many
authors propose a variety of models to describe the evolution of
our universe, like the modified gravity \cite{xh} for example.
Among these and opposite to the extending Hilbert-Einstein action
for general relativity modifications, there exist a class of
models that are based on searching for a proper equation of
state(EoS) for the matter-energy fluid. Initially, this class of
models are exploited in the context of the perfect fluid. The
viscosity concept is introduced into dark energy study relatively
lately. And now it seems to play a more and more important and
practical role in the more realistic cosmology model
constructions. 
Other earlier attempts in this line can be found in
references\cite{CWM,SN,IB,IB1,IB2,TP}. Additionally, for more
details Gr{\o}n have given a very useful review for the subject in
reference\cite{G}.

Viscosity  is a concept in fluid mechanics related to velocity
gradient and is divided into two classes, shear viscosity and bulk
viscosity. In viscous cosmology, shear viscosity comes into play
in connection with spacetime  anisotropy. An analytic formula for
the traceless part of the anisotropy stress tensor has been
derived by S.Weinberg\cite{W}. Meanwhile, a bulk viscosity usually
functions in an isotropic universe. Under the
Friedmann-Robertson-Walker (FRW) framework, the energy-momentum
tensor at most has a bulk viscosity term as
$T_{ik}=(p-\zeta\theta)g_{ik}$($\zeta$ is bulk viscosity, and
$\theta$ is the expansion scalar). Additionally, bulk viscosity
related to the grand-unified-theory phase transition \cite{PL} may
lead to explain the cosmic acceleration expansion.

At present, a large number of models exist but without very
effective methods either verifying them or ruling them out. For
this reason, there is a strong need for diagnostic techniques. The
statefinder diagnostic pair $\{r,s\}$, which is purely geometric
quantities introduced by Sahni, Saini, Starobinsky and
Alam\cite{VS}, provides us a very useful method to discriminate
cosmological models. The statefinder pair probes the expansion
dynamics of the universe through higher derivatives of the
expansion factor $\dddot{a}$($a$ is the scale factor) which is a
natural companion to the deceleration parameter that depends upon
$\ddot{a}$. Different models on the $s-r$ plane accordingly show
different trajectories. For example, the spatially flat
$\Lambda$CDM scenario corresponds to a fixed point $\{0,1\}$ in
the statefinder diagnostic pair $\{r,s\}$ plane, with which the
`distance' of other Dark Energy (DE) models from $\Lambda$CDM can
therefore be established on the $s-r$ plane \cite{UA}.
Additionally, the statefinder pair has possessed the merits that
can discriminate among a large amount of models including
$\Lambda$CDM, quintessence, kinessence, Chaplygin gas(see
references \cite{VS,UA,VG,WZ}). With the introduction of new
observational techniques and increasing improvement of
measurement, it is certainly for us to get more precision data and
richer information about the geometric quantities, with
possibilities to evaluate more practical cosmology models, like
those reasonable tries by considering viscosity to abandon the
commonly used and simplest perfect fluid approximation to real
cosmic media, and at that time some models would be either
verified or ruled out by the statefinder diagnostic pair and other
astrophysics observations. More reference about recent and future
experiments can be found in the review paper\cite{SH}.

This paper is arranged as follows: In Sec.\ref{sec:General
formalism}, the general formalism is presented for following
discussions; In Sec.\ref{sec:cosmology with constant bulk
viscosity}, we discuss the difference between perfect fluid models
and non-perfect fluid models from the viewpoint of density and
trajectories of the diagnostic pair$\{r,s\}$, especially in
Quiessence model\cite{UA,VG,qm}; In Sec.\ref{sec:Kinessence model
with general bulk viscosity}, our attentions are focused on
variable bulk viscosity dark energy model with general EoS. The
entropy of system is deeply discussed there apart from statefinder
pair diagnostics, as we have realized that the thermodynamics can
reflect more important global characters for a complicated system,
like our observable universe as shown by Brevik, Nojiri,
Odintsov and Vanzoe in the reference \cite{IB2}. And the last
section is devoted to our conclusions.

\section{General formalism}         \label{sec:General formalism}
 Now, we introduce the basic framework for our discussions. That is, in
 Friedmann-Robertson-
 Walker(FRW)
 cosmology the metric of the system is chosen as:
\begin{equation}
ds^{2}=-dt^{2}+a^{2}(\frac{dr^{2}}{1-kr^{2}}+r^{2}d\Omega^{2})
\end{equation}
where $a$, and $k$ are the scale factor and space curvature,
respectively. The Einstein equations take the usual form
\begin{equation}\label{eq:G}
G_{\mu\nu}=R_{\mu\nu}-\frac{1}{2}Rg_{\mu\nu}=8\pi GT_{\mu\nu}
\end{equation}
Note that we have included the cosmological constant $\Lambda$ in
the energy-momentum tensor $T_{\mu\nu}$. In the FRW cosmology with
bulk viscosity, the stress-energy-momentum tensor is written as:
\begin{equation}
T_{\mu\nu}=(\rho+p)U_{\mu}U_{\nu}+pg_{\mu\nu}-\zeta\theta h_{\mu\nu}
\end{equation}
where $\zeta$ is the bulk viscosity, $\theta$ the expansion factor
defined by $\theta=U^{\mu}_{;\mu}=3\dot{a}/a$, and the projection
tensor $h_{\mu\nu}$ is defined by
$h_{\mu\nu}=g_{\mu\nu}+U_{\mu}U_{\nu}$ with $U^{\mu}$ being the four
velocity of the fluid on the comoving coordinates. On the
thermodynamical grounds, $\zeta$ is conventionally chosen to be a
positive quantity and may depend on the cosmic time $t$, or the
scale factor $a$, or the energy density $\rho$, etc. Through a
series of calculations, the non-vanishing equations in
Eq.(\ref{eq:G}) are
\begin{eqnarray}
\frac{\dot{a}^{2}}{a^{2}}&=&\rho  \label{eq:Einstein}\\
\dot{\rho}&=&-3H(\rho+\widetilde{p})\label{eq:conservation}
\end{eqnarray}
where $\widetilde{p}$ is an equivalent pressure defined by
$\widetilde{p}=p-\zeta\theta$ and $H$ denotes the Hubble parameter.
Additionally, we use the unity convention $8\pi G/3=c=1$. Normally,
if we have known the information about both the equation of
state(EoS) and the bulk viscosity $\zeta$, the fate of the universe
would have been determined by the Eqs.(\ref{eq:Einstein}) and
(\ref{eq:conservation}).

In the following parts, the viscous cosmology will be checked with
the use of statefinder diagnostic pair $\{r,s\}$ parameters. Here
we first present their general expressions explicitly. The
statefinder pair $\{r,s\}$ is defined by(see reference \cite{VS})
\begin{eqnarray}                \label{eq:rs}
r=\frac{\dddot{a}}{aH^{3}}, \quad s=\frac{r-1}{3(q-\frac{1}{2})}
\end{eqnarray}
where $q=-\ddot{a}/aH^{2}$ is the deceleration parameter. Combined
with Eqs.(\ref{eq:Einstein}) and (\ref{eq:conservation}),
Eq.(\ref{eq:rs}) becomes:
\begin{align}
q&=\frac{3}{2}\frac{p}{\rho}-\frac{9}{2}\frac{\zeta}{\sqrt{\rho}}+\frac{1}{2}\label{eq:Q}\\
r&=-\frac{3}{2}\left[\frac{\dot{p}}{\rho^{\frac{3}{2}}}-3\frac{\dot{\zeta}}{\rho}+3(q+1)\frac{\zeta}{\sqrt{\rho}}\right]+1
\tag{\ref{eq:Q}$a$}\\
s&=\frac{3\dot{\zeta}\sqrt{\rho}-\dot{p}-3(q+1)\zeta\rho}{3(\sqrt{\rho}p-3\rho\zeta)}\tag{\ref{eq:Q}$b$}
\end{align}
From the above formulas, we can see that the diagnostic statefinder
pair $r$ and $s$ are related to the quantities $\rho$, $p$,
$\dot{p}$, $\zeta$ and $\dot{\zeta}$, among which $\rho$, $p$,
$\dot{p}$ can be reduced to one quantity if we have known the EoS.
And for the little known global quantities $\zeta$ and $\dot{\zeta}$
in the viscous cosmology, we can discuss the models with the
simplest case of $\zeta=\zeta_{0}$(constant) first as shown in the
next section
Recently, some authors have proposed a few fresh opinions about
the possible forms of bulk viscosity $\zeta$ such as
$\zeta=\tau\theta$ for discussing dark energy cosmology and dark
fluid properties (see reference \cite{IB,XM}), which has builded
up a relationship between the bulk viscosity $\zeta$ and the scale
factor $a$. We will discuss such models with variable bulk
viscosity in the section \ref{sec:Kinessence model with general
bulk viscosity}.

\section{cosmology with constant bulk viscosity}\label{sec:cosmology with constant bulk viscosity}
In this section, we treat the universe model with a more realistic
situation as containing two main media parts , that is, one is the
mainly non-relative matter component $M$ while  another is the
dominated dark energy $X$ component. Thus, the total pressure and
density can be expressed as
\begin{eqnarray}
p=p_{M}+p_{X}\simeq p_{X},\quad \rho=\rho_{M}+\rho_{X}
\end{eqnarray}
where the pressure of matter $p_{M}$ is a negligible quantity. And
the EoS of the $X$ part can take a usual factorization form
$p_{X}=w\rho_{X}$($w$ is called state parameter).

From the viewpoint of the bulk viscosity, the simplest case is
thought to be a constant bulk viscosity $\zeta=\zeta_{0}$. In this
section we mainly discuss the statefinder diagnostic pairs to two
well known cosmological dark energy models, added with a constant
bulk viscosity. Through the comparisons of viscous and non-viscous
models, it is beneficial for us to understand the role of cosmic
viscosity, the properties of the cosmic models with common EoS and
further our physical universe more comprehensively.

\subsection{$\Lambda$CDM model}\label{subsec:cdm model}
So far as we know, the $\Lambda$CDM model, with mainly two
components: the cosmological constant and cold dark matter, may be
the simplest and the most consistent one with observation data.
And also it has been deeply studied with no viscosity assumption.
In this subsection, we will continue to study it in the context of
the viscous cosmology by using the statefinder diagnostic pair.

\begin{figure}
  \includegraphics{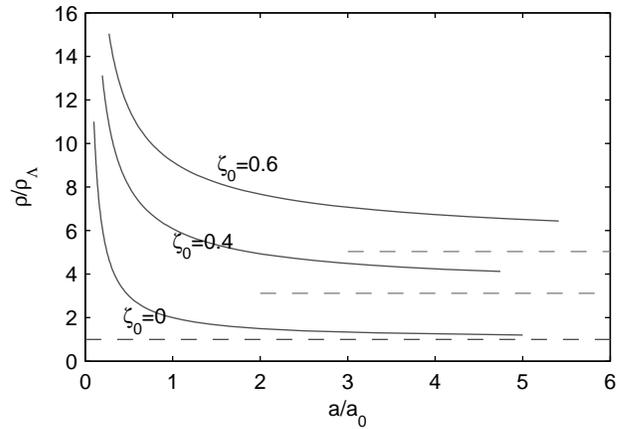}\\
  \caption{On the $a-\rho$ plane, we can see the differences at the different values of
  $\zeta_{0}$.
   The horizontal lines represent the values, to which the next higher curves approach. Here we
  assume for simplicity
   the parameters($\rho_{\Lambda}=1$ and $B=1$). Note that $\zeta_{0}=0$ means without viscosity.}\label{fig:statefinder1}
\end{figure}

Considering the Einstein equation (\ref{eq:Einstein}) and energy
conservation equation (\ref{eq:conservation}), the following
integration is obvious:
\begin{equation}\label{eq:integration1}
\int\frac{d\rho_{M}}{\rho_{M}-3\zeta_{0}\sqrt{\rho_{M}+\rho_{\Lambda}}}
=-3\int\frac{da}{a}
\end{equation}
where $\rho_{X}$ has become $\rho_{\Lambda}$ (the vacuum energy
density). By working out the above integration, we get the relation
between $\rho$ and $a$:
\begin{equation}\label{eq:pa}
(\rho-3\zeta_{0}
\sqrt{\rho}-\rho_{\Lambda})\left(\frac{2\sqrt{\rho}-3\zeta_{0}-\sqrt{4\rho_{\Lambda}+9\zeta_{0}^{2}}}{2\sqrt{\rho}-3\zeta_{0}+\sqrt{4\rho_{\Lambda}+9\zeta_{0}^{2}}}\right)^{\xi}=\frac{B}{a^{3}}
\nonumber
\end{equation}
where the $B$ is an integration constant and the $\xi$ is a constant
defined by
\begin{equation*}
    \xi=\frac{3\zeta_{0}}{\sqrt{4\rho_{\Lambda}+9\zeta_{0}^{2}}}
\end{equation*}
Figure \ref{fig:statefinder1} demonstrates the $a(t))-\rho$ relation
with the different values of $\zeta_{0}$. There is a minimum value
of the $\rho$ denoted by $\rho_{\Lambda}^{eff}$ called the
effective vacuum energy density(EVED) corresponding to
$a\rightarrow\infty$ . The expression of EVED is as
\begin{equation}\label{eq:rho_m}
    \rho_{\Lambda}^{eff}=\frac{1}{4}\left(3\zeta_{0}+\sqrt{4\rho_{\Lambda}+9\zeta_{0}^{2}}\right)^{2}.
\end{equation}
The equation (\ref{eq:rho_m}) has such a limit clearly:
\begin{equation*}
    \lim_{\zeta_{0}\rightarrow0}\rho_{\Lambda}^{eff}=\rho_{\Lambda}
\end{equation*}
which returns to the non-viscosity situation directly. Since
$\rho_{\Lambda}$ denotes the vacuum energy density under the
non-viscosity case, we might as well call it the conventional
vacuum energy density(CVED) in contrast to the EVED.

For the simplicity to discuss the trajectories on the figure
\ref{fig:statefinder1} which corresponds to the different values
of the $\zeta_{0}$, we apply the relative changing rate as
$\varsigma$ defined by
\begin{equation}\label{eq:change_rate}
    \varsigma=\left|\frac{\triangle\rho}{\rho(a)}\right|=\frac{\left|\rho(a+\triangle a)-\rho(a)\right|}{\rho(a)}
\end{equation}
where the $\triangle a$ denotes a small change of the scale factor
$a(t)$. After the interference of the bulk viscosity, the
densities as a whole increase much more while the $\triangle\rho$
changes a little from numerical analysis and the figure
\ref{fig:statefinder1}, and then the $\varsigma$ becomes smaller
than before. And the trend of the changes is: the larger the bulk
viscosity($\zeta_{0}$), the bigger the value of EVED and the
smaller the $\varsigma$. 
We may sentence that the bulk viscosity stabilizes the evolution
of the density and blocks then the rapid change of the universe,
which is easily understandable by physics intuition as in the
friction case that the friction force hinders dynamically the rapidly kinematic
movements.

Under this model, we put $p=p_{M}+p_{\Lambda}$,
$\rho=\rho_{M}+\rho_{\Lambda}$ into Eq.(\ref{eq:Q}) and get the
statefinder pair as:
\begin{align}
q&=-\frac{3}{2}\Omega_{\Lambda}-\frac{9}{2}\frac{\zeta_{0}}{H}+\frac{1}{2} \label{eq:a}\\
r&=\frac{27}{4}\left[3\frac{\zeta_{0}^{2}}{H^{2}}-(1-\Omega_{\Lambda})\frac{\zeta_{0}}{H}\right]+1\tag{\ref{eq:a}$
b$}\\
s&=-\frac{3}{2}\frac{\zeta_{0}}{H}\left(1-\frac{1}{\Omega_{\Lambda}+3\frac{\zeta_{0}}{H}}\right)\tag{\ref{eq:a}$
c$}
\end{align}
where the $\Omega_{\Lambda}$ is the vacuum density parameter defined
by $\Omega_{\Lambda}=\rho_{\Lambda}/H^{2}$.

\begin{figure}
  \includegraphics{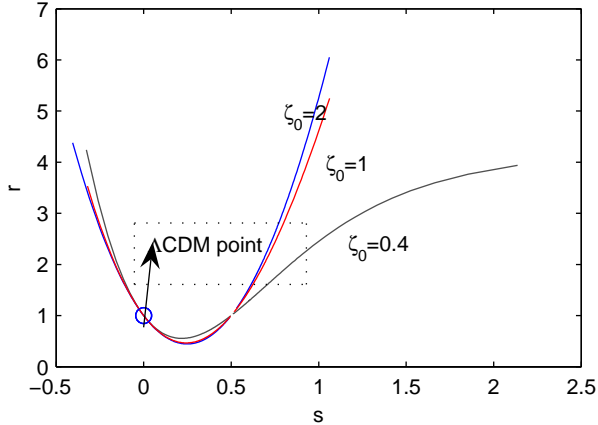}\\
  \caption{On the $s-r$ plane, we can see the differences with the different values of $\zeta_{0}$
  . When $\zeta_{0}$ takes a smaller value, the trajectory looks not similar to a parabola.
   However, as $\zeta_{0}$ becomes bigger and bigger, its trajectory tends like the parabola  more and more.  Here we
   assume
   the parameter($\rho_{\Lambda}=1$) for simplicity. Note that $\zeta_{0}=0$(a circle) means no viscosity in this case.}
   \label{fig:statefinder2}
\end{figure}

 The expressions (\ref{eq:a}) demonstrate that the diagnostic pair $\{r,s\}$ is dependent on a single dynamically changing
quantity $H$. We draw on the figure \ref{fig:statefinder2} the
trajectories of the statefinder pair $\{r,s\}$. The point
$\{0,1\}$ corresponds to the diagnostic pair $\{r,s\}$ for the $\Lambda$CDM model
with no viscosity, and the curves which with the larger bulk
viscosity $\zeta_{0}$ are more parabola-like represent viscous
situations. The $\Lambda$CDM models that either are viscous or
non-viscous are therefore differentiated by the trajectories on
the $s-r$ parameters plane.

\subsection{Quiessence model}\label{subsec:Quiessence model}
The Quiessence model which characterizes itself with the EoS
:\[p_{X}=w_{0}\rho_{X}\]($w_{0}$ is constant, but asked not -1 as
reason shown below, with contrast to the cosmological constant case)
has been used recently to describe the dark energy behaviors. It is
interesting to consider such model under the viscous situation.
After assuming no interaction between two fluids $M$ and $X$, we can
decompose them , that is, the $X$ and $M$ satisfy the
Eqs.(\ref{eq:Einstein}) and (\ref{eq:conservation}), respectively.
By writing the $X$ part out independently, we have:
\begin{equation}\label{eq:integration2}
    \int\frac{d\rho_{X}}{(1+w_{0})\rho_{X}-3\zeta_{0}\sqrt{\rho_{X}}}=-3\int\frac{da}{a}
\end{equation}
By working the above integration out, we get
\begin{equation}\label{eq:pXa}
    \rho_{X}=\left[\frac{3\zeta_{0}}{1+w_{0}}+B(1+z)^{\frac{3(1+w_{0})}{2}}\right]^{2}
\end{equation}
where the redshift is denoted by $z$ with $z=a_{0}/a-1$ and $B$ is
an integration constant. And the density of the non-relative
matter $M$ component has possessed the following scaling relation
from Eqs.(\ref{eq:Einstein}) and (\ref{eq:conservation})
\[\rho_{M}=C(1+z)^{3}\] with a positive integration constant $C$.
Then the value of the total energy density as an additive quantity
can be easily composed as:
\begin{equation}\label{eq:quiessence_rho}
    \rho=\left[\frac{3\zeta_{0}}{1+w_{0}}+B(1+z)^{\frac{3(1+w_{0})}{2}}\right]^{2}+C(1+z)^{3}
\end{equation}
\begin{figure}
  \includegraphics{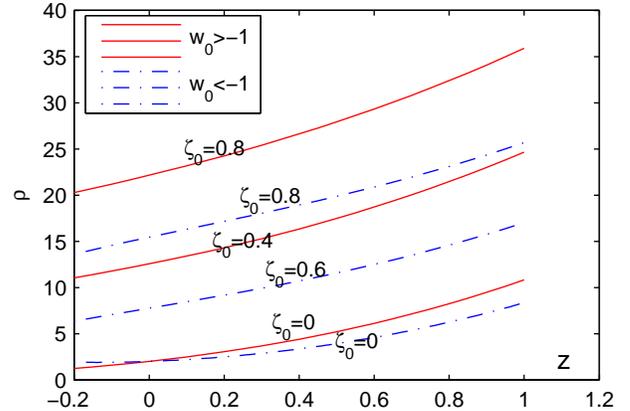}\\
  \caption{In the $\rho-z$ plane, we can see the differences at the different values of $\zeta_{0}$
  . The viscosity makes the density bigger than that of no-viscosity cases. Here we
   assume
   the parameter($B=C=1$, and $w_{0}=-0.5 or -1.5$).Note that $\zeta_{0}=0$ means that there is no viscosity.}
   \label{fig:statefinder3}
\end{figure}
It is a function about the scaling relation with variable $z$.
Moreover, the concept of relative changing rate parameter
$\varsigma$ will be used again to investigate the effects of the
bulk viscosity for clarifications.

For the simplicity of discussing the parameter $\varsigma$, we
first draw the relation of the Eq.(\ref{eq:quiessence_rho}) on the
figure \ref{fig:statefinder3}. The definition of the $\varsigma$
is modified in this case as \[\varsigma=\frac{|\rho(z+\triangle
z)-\rho(z)|}{\rho(z)}\] where the $\triangle z$ denotes a small
change of the redshift $z$. With the same analysis as in the above
subsection \ref{subsec:cdm model}, we can still obtain the
conclusion that the bulk viscosity stabilizes the evolution of the
density in this model. It is worth noting that on the figure
\ref{fig:statefinder3} the trajectories are divided into two
classes corresponding to  $w_{0}>-1$ and $w_{0}<-1$ respectively,
and $w_{0}=-1$ is correspondence to the singularity as shown in
Eq.(15) clearly. The phantom dividing phenomenon\cite{ml} can also
appear on the statefinder pair plane as illustrated in the following
discussions.

The statefinder diagnostic pair of the Quiessence model can be
gotten, when we put $\rho=\rho_{M}+\rho_{X}$ and $p\simeq
w_{0}\rho_{X}$ into Eq.(\ref{eq:Q}). The results are:
\begin{align}
q&=\frac{3}{2}w_{0}\Omega_{X}-\frac{9}{2}\frac{\zeta_{0}}{H}+\frac{1}{2}\label{eq:1} \\
r&=\frac{9}{2}(1+w_{0})w_{0}\Omega_{X}-\frac{27}{4}\frac{\zeta_{0}}{H}(w_{0}\Omega_{X}-3\frac{\zeta_{0}}{H}+1)+1
\tag{\ref{eq:1}$a$}\\
s&=\frac{(1+2w_{0})w_{0}\Omega_{X}}{2(w_{0}\Omega_{X}-3\frac
{\zeta_{0}}{H})}-\frac{3}{2}\frac{\zeta_{0}}{H}+\frac{1}{2}\tag{\ref{eq:1}$b$}
\end{align}
where the $\Omega_{X}$ denotes the dark energy density parameter
defined by $\Omega_{X}=\rho_{X}/H^{2}$. Considering
Eqs.(\ref{eq:Einstein}) and (\ref{eq:quiessence_rho}), we can
ultimately transform $q, r, s$ into such quantities depending on
the redshift $z$ only.

\begin{figure*}
  \includegraphics{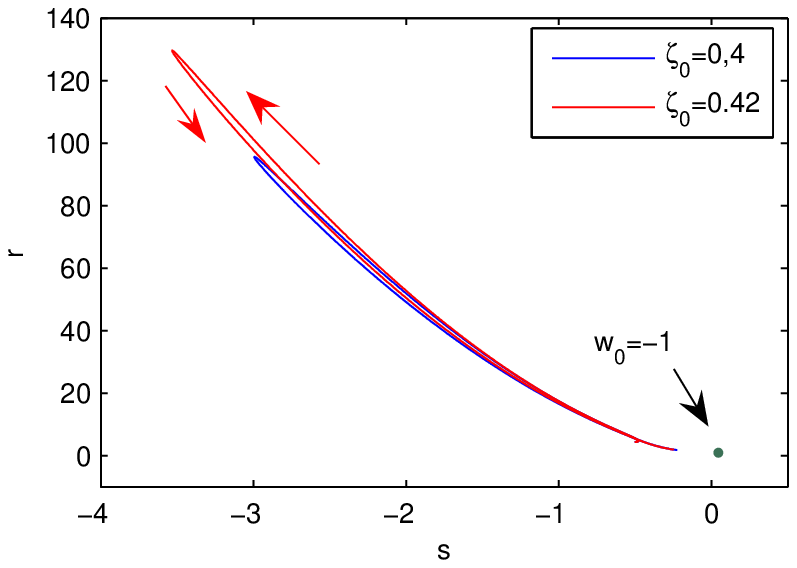}
  \includegraphics{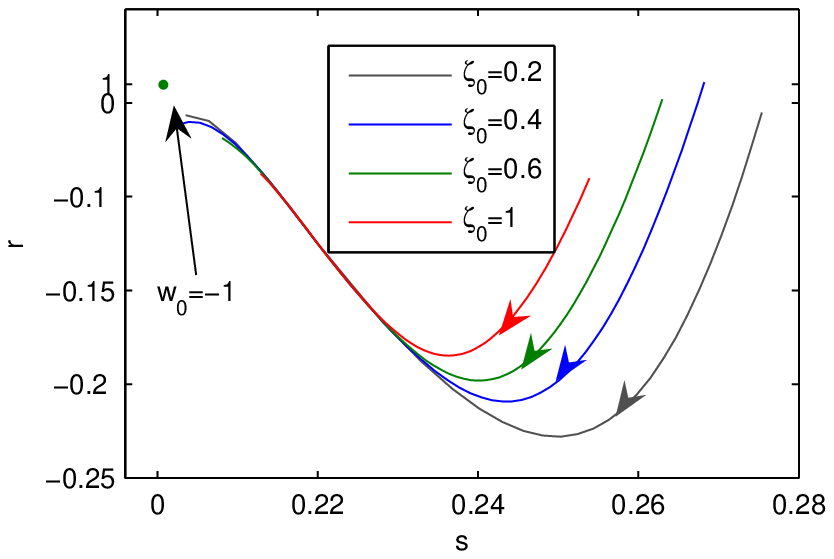}
  \includegraphics{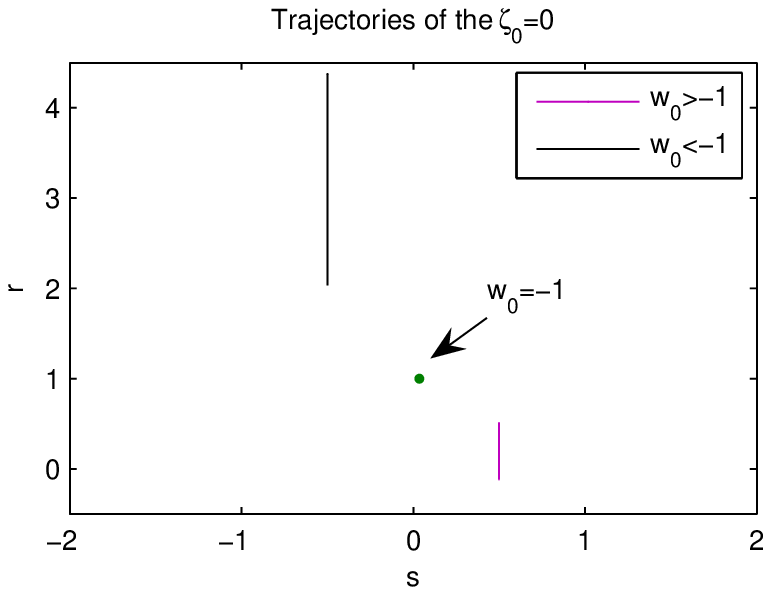}
\caption{The above three panels are numbered from up to down as
$4.1$,$4.2$ and $4.3$: $4.1$ is at the situation of $w_{0}=-1.5$;
while $4.2$ is in the case of $w_{0}=-0.5$ and $4.3$ is for
viscosity free $\zeta_{0}=0$ condition. Arrows represent the
directions of the evolutions of statefinder diagnostic pair about
time. Here we have assumed for simplicity the parameters ($B=C=1$)
in the Eq.(\ref{eq:quiessence_rho}).}
   \label{fig:statefinder4}
\end{figure*}

 The situation of the $w_{0}=-1$ 
 is thought as the border-case called `phantom divide' (see reference \cite{WH}).
 The singular property of the `phantom divide' can also be described by
 the trajectories on the statefinder pair plane in the figure
 \ref{fig:statefinder4}. The corresponding relationship between the
 trajectories on the figure \ref{fig:statefinder4} and the values of
 the parameters are arranged in the table \ref{tab:table1}.

\begin{table}
\caption{\label{tab:table1}Relationship between Figures and
parameters}
\begin{ruledtabular}
\begin{tabular}{lccr}
&$w_{0}>-1$&$w_{0}=-1$&$w_{0}<-1$\\
\hline $\zeta_{0}>0$&curves on $4.2$\footnotemark[2] &
bullets\footnotemark[0] on $4.1$\footnotemark[1],$4.2$\footnotemark[2] & curves on $4.1$\footnotemark[1]\\
$\zeta_{0}=0$&curves on $4.3$\footnotemark[3]& bullets on $4.3$\footnotemark[3] & curves on $4.3$\footnotemark[3]\\

\end{tabular}
\end{ruledtabular}
\footnotemark[0]{bullet hereafter refers to point for diagnostic
pair $(s,r)=(0,1)$, i.e., corresponding to $\Lambda$CDM model}\\
\footnotemark[1]{$4.1$ denotes the first panel in the
figure\ref{fig:statefinder4}.}\\
\footnotemark[2]{$4.2$ refers to the
second panel in the figure \ref{fig:statefinder4}.}\\
\footnotemark[3]{$4.3$ represents the third panel in the figure
\ref{fig:statefinder4}.}
\end{table}

 Obviously the point of $(0,1)$ that denotes the
situation of the $w_{0}=-1$ divides the trajectories into two
parts, the curves of Fig.$4.1$ and the curves of Fig.$4.3$. The
directions of the evolutions demonstrate that our universe is
approaching the state of the $w_{0}=-1$ and then keeps itself
stable there, which is consistent with results from lots of data
analysis and cosmic simulations as favored to the $\Lambda$CDM
model for describing later stage universe evolutions.

\section{DE model with variable bulk viscosity and EoS, the entropy}\label{sec:Kinessence model with general bulk viscosity}

In this section, we mainly discuss the viscous dark energy model
which characterizes itself with the variable state parameter $w$.
We here take the form of the bulk viscosity as
$\zeta=\zeta_{1}\dot{a}/a$(see reference \cite{IB,XM}) which is a
particular situation of the general approaches as proposed in the
reference \cite{SC}. Some thermodynamic discussions, especially
the entropy expression (an instructive way to obtain the entropy
expression for viscous cosmology from generalized Cardy-Verlinde
entropy formula can be found in Ref.\cite{bo}), are made first,
and then the trajectories of the model on the statefinder
diagnostic pair plane are shown. For simplicity, we only discuss
the universe filled by the fluid of only one component with the
following EoS \cite{SW}
\begin{equation}\label{eq:pq}
p=(\gamma-1)\rho+p_{1}
\end{equation}
where the $p_{1}$ is a constant (as subscript 0 often indicating
present value). For system only consisting of same mass particles,
like baryons, its pressure is proportional to $nm$ where $n$ is
the number density with $m$ as the mass for one particle. The
$\gamma$ is the adiabatic exponent or barotropic factor,
specifically, $\gamma=5/3$ in the case of extreme non-relativity,
and $\gamma=4/3$ in the case of extreme relativity. As for other
general cases, $\gamma$ will take complicated forms.

 The $\gamma-1$ will be equivalent
to the state parameter $w$ (hereafter) if the absolute value of
$p_{1}$ is much smaller than that of pressure $p$. The bound on
the state parameter $w$
is given out as (see reference \cite{AM,TRC,RO})
\[-1.38<w<-0.8\]  and thus the adiabatic exponent $\gamma$ gets its own
constraint as $-0.38<\gamma<0.2$.

By directly calculations from the Eqs.(\ref{eq:Einstein}) and
(\ref{eq:conservation}), the scale factor is obtained as:

Provided $\widetilde{\gamma}<0$,
\begin{equation}\label{eq:a_1}
a(t)=a_{0}\left[\cosh\left(\frac{t-t_{0}}{2T_{1}}\right)+\widetilde{\gamma}\theta_{0}T_{1}\sinh\left(\frac{t-t_{0}}{2T_{1}}\right)\right]^{\frac{2}{3\widetilde{\gamma}}}
\end{equation}

If $\widetilde{\gamma}>0$,
\begin{equation}\label{eq:a_2}
a(t)=a_{0}\left[\cos\left(\frac{t-t_{0}}{2T_{2}}\right)+\theta_{0}\widetilde{\gamma}T_{2}\sin\left(\frac{t-t_{0}}{2T_{2}}\right)\right]^{\frac{2}{3\widetilde{\gamma}}}
\end{equation}

where $T_{1}$,and $T_{2}$ are defined by
\begin{eqnarray*}
    T_{1}=\frac{1}{3\sqrt{-\widetilde{\gamma}p_{1}}},\quad
    T_{2}=\frac{1}{3\sqrt{\widetilde{\gamma}p_{1}}}
\end{eqnarray*}
And where
\begin{equation*}
    \widetilde{\gamma}=\gamma-3\zeta_{1}
\end{equation*}
is an effective parameter for the $\gamma$, and the $\rho_{0}$,
$\theta_{0}$, $t_{0}$ are the present energy density, expansion
scalar, and cosmic time, respectively.

\begin{figure}
  \includegraphics{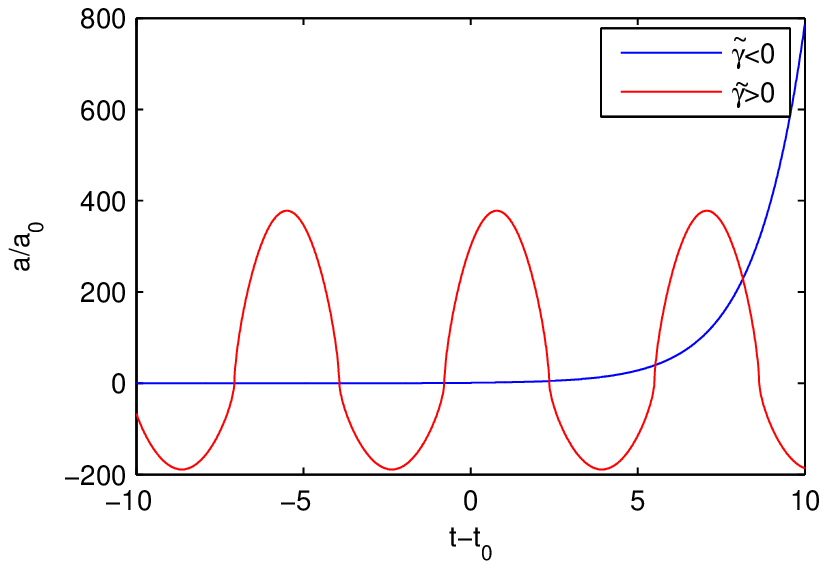}\\
  \caption{On the $t$-$a$ plane, the scale factor $a(t)$ at the case of $\widetilde{\gamma}>0$
shows an oscillating expansion cosmology model while the
$\widetilde{\gamma}<0$ case corresponds to the late time
accelerating expansion universe
}\label{fig:statefinder12}
\end{figure}

Eqs.(\ref{eq:a_1}) and (\ref{eq:a_2}) are compared by the use of
curves on the figure \ref{fig:statefinder12}. As for the case of
$\widetilde{\gamma}<0$, it can successfully explain the cosmic
accelerating expansion in the late evolution universe. Conversely,
it is impossible to do so for the case of $\widetilde{\gamma}>0$,
partly because there are negative values of the scale factor $a$
which seem un-physical by the direct mathematical treatment. So it
is proper for us to only consider the case $\widetilde{\gamma}<0$
in the following discussion.

The number of effective parameters from (\ref{eq:a_1}) is reduced
to three: $T_{1}$, $\theta_{0}$, and $\widetilde{\gamma}$. For the
convenience of following discussions, it is beneficial here to
consider possible constraint to these parameters  from the
favorable fact $\dot{a}>0$ particularly in the late universe
evolution. By calculating $\dot{a}$ with the use of
Eq.(\ref{eq:a_1}), we can have
\begin{equation}\label{eq:a_dot}
    \dot{a}=\frac{a}{3\widetilde{\gamma}T_{1}}x^{3\widetilde{\gamma}/2}\left[\sinh(\frac{t-t_0}{2T_{1}})+W\cosh(\frac{t-t_{0}}{2T_{1}})\right]
\end{equation}
where $W$ is defined by $W=\widetilde{\gamma}\theta_{0}T_{1}$ and
$x$ is defined by $x=a_{0}/a=1+z$($z$ denotes the redshift). We have
known that $a>0$, $\widetilde{\gamma}<0$, and $T_{1}>0$, and thus
$\dot{a}>0$ is equivalent to
\begin{equation}\label{eq:par_ran}
    W<\frac{1-\exp(\frac{t-t_{0}}{T_{1}})}{1+\exp(\frac{t-t_{0}}{T_{1}})}
\end{equation}
Corresponding to $t\rightarrow\infty$, as our interested later
stage evolution for the observable universe, the right hand of the
above inequality has possessed a limit as following
\begin{equation*}
    -1<\frac{1-\exp(\frac{t-t_{0}}{T_{1}})}{1+\exp(\frac{t-t_{0}}{T_{1}})}
\end{equation*}
So the constraint (\ref{eq:par_ran}) of $W$ becomes
\begin{equation*}
    W\leq-1
\end{equation*}

Here we also give out, respectively, the expressions of density
and expansion factor depending on the relevant variables $t$ and
$z$, which will be used in our entropy expression calculations:
\begin{eqnarray}
  \rho(z)&=&\rho_{0}\left[(1+\frac{p_{1}}{\rho_{0}\widetilde{\gamma}})(1+z)^{3\widetilde{\gamma}}-\frac{p_{1}}{\rho_{0}\widetilde{\gamma}}\right]\label{eq:rho_W}\\
 \theta(t) &=& \frac{\theta_{0}}{W}\cdot\frac{\sinh(\frac{t-t_{0}}{2T_{1}})+W\cosh(\frac{t-t_{0}}{2T_{1}})}{\cosh(\frac{t-t_{0}}{2T_{1}})+W\sinh(\frac{t-t_{0}}{2T_{1}})}\nonumber \\
  \theta(z) &=&-\frac{\theta_{0}}{W}\cdot\sqrt{1+(W^{2}-1)(1+z)^{3\widetilde{\gamma}}}\label{eq:exp_x}
\end{eqnarray}

For the perfect fluid models in a closed cosmic system, the cosmic
media is regarded without dissipation and the entropy is a
conservation quantity with $dS/dt=0$($S$ denotes the entropy of
the system per unit volume). However, considering non-perfect
fluid models, the entropy will change. Now we turn our attentions
on the entropy of the model as introduced in this section.

The relevant general formulas to be employed (see reference
\cite{IB1,SW1,AHT}) are:
\begin{equation}\label{Eq:entropy vector}
    S^{\mu}_{;\mu}=\frac{2\eta}{T}\sigma_{\mu\nu}\sigma^{\mu\nu}+\frac{\zeta}{T}\theta^{2}+\frac{1}{\kappa T^{2}}Q_{\mu}Q^{\mu}
\end{equation}
where the $S^{\mu}$ is the entropy four-vector,  $\eta$ the shear
viscosity, $T$ the temperature,  $\zeta$  the bulk viscosity,
$\sigma_{\mu\nu}$ the shear tensor, $\theta$  the expansion
factor, $\kappa$ the thermal conductivity and $Q_{\mu}$ as the
space-like heat flux density four-vector.

The entropy four-vector $S^{\mu}$ is defined by
\begin{equation}\label{Eq:entropy}
    S^{\mu}=n\sigma U^{\mu}+\frac{1}{T}Q^{\mu}
\end{equation}
where the $n\sigma$ is the ordinary entropy per unit volume($n$
denoted the particle number per unit volume with $\sigma$ as the
entropy of one particle).
 The expansion tensor $\theta_{\mu\nu}$ is
defined as:
\begin{equation*}
    \theta_{\mu\nu}=\frac{1}{2}(U_{\mu;\alpha}h^{\alpha}_{\nu}+U_{\nu;\alpha}h^{\alpha}_{\mu})
\end{equation*}
The scalar expansion factor is
$\theta=\theta^{\mu}_{\mu}=U^{\mu}_{;\mu}$. The shear tensor is
defined as
\begin{equation*}
    \sigma_{\mu\nu}=\theta_{\mu\nu}-\frac{1}{3}h_{\mu\nu}\theta
\end{equation*}
which is traceless, that is $\sigma_{\mu}^{\mu}=0$ and where the
$h_{\mu\nu}$ has been defined by
$h_{\mu\nu}=g_{\mu\nu}+U_{\mu}U_{\nu}$. Defining the
four-acceleration of the fluid as
$A_{\mu}=\dot{U}_{\mu}=U^{\nu}U_{\mu;\nu}$, the space-like heat
flux density four-vector is given by
\begin{equation*}
    Q^{\mu}=-\kappa h^{\mu\nu}(T_{,\nu}+TA_{\nu})
\end{equation*}
In the case of thermal equilibrium, $Q^{\mu}=0$.

Under the background of FRW metric, we can have
\begin{eqnarray*}
  \sigma_{\mu\nu} = 0, \quad
  \theta=3H
\end{eqnarray*}
Eqs.(\ref{Eq:entropy vector}) and (\ref{Eq:entropy}) yield
\begin{eqnarray*}
  S^{\mu}_{;\mu} = \frac{\zeta_{1}}{3T}\theta^{3},\quad
  S^{0}= n\sigma,\quad
  S^{i}=0(i=1,2,3.)
\end{eqnarray*}
After taking account of FRW metric, we can get the following
differential equation as 
\begin{eqnarray}
    S^{0}_{,0}+\theta S^{0}=\frac{\zeta_{1}}{3T}\theta^{3}\label{Eq:differential functon}
\end{eqnarray}
where the `$,0$' denotes time derivative $d/dt$. We assume that
the fluctuation of temperature is so small that it is negligible.
The Eq.(\ref{Eq:differential functon}) can be transformed as
\begin{equation}\label{eq:S_theta}
    \frac{\theta_{0}}{2T_{1}W}\left(1-W^{2}\frac{\theta^{2}}{\theta_{0}^{2}}\right)\frac{dS^{0}}{d\theta}+\theta
    S^{0}=\frac{\zeta_{1}}{3T}\theta^{3}
\end{equation}

Then, based on Eq.(\ref{eq:a_1}), we work the differential
Eq.(\ref{Eq:differential functon}) out and get:
\begin{equation}\label{eq:entropy zero}
    S^{0}=\frac{1}{1-W^{2}\theta^{2}/\theta_{0}^{2}}\left(\frac{\zeta_{1}T_{1}W}{6\theta_{0}T}\theta^{4}+Ce^{\frac{T_{1}\theta_{0}}{W}}\right)
\end{equation}
where the $C$ is an integral constant. From the equation of
(\ref{eq:entropy zero}), we find that when $\theta=\theta_{0}/W$,
the $S^{0}$ approaches to infinity if we did not consider the
values of the constant $C$. To avoid this un-physically
mathematical phenomenon, we choose the value of $C$
as\[C=-\frac{\zeta_{1}T_{1}\theta_{0}^{3}}{6TW^{3}}e^{-\frac{T_{1}\theta_{0}}{W}}\]
which will lead to such a limit that
\[\lim_{\theta\rightarrow\theta_{0}/W}S^{0}=-\frac{\zeta_{1}T_{1}\theta_{0}}{3TW}\theta^{2}\]
without the formal singularity. The expression (\ref{eq:entropy
zero}) of the $S^{0}$ therefore becomes:
\begin{equation}\label{eq:entropy zero1}
    S^{0}(\theta)=-\frac{\zeta_{1}T_{1}\theta_{0}}{6W^{3}T}\cdot\frac{W^{4}\theta^{4}-\theta_{0}^{4}}{W^{2}\theta^{2}-\theta_{0}^{2}}
    \end{equation}
    or in a more easily reading form
\begin{equation}\label{eq:entropy zero2}
    S^{0}(z)=-\frac{\zeta_{1}T_{1}\theta_{0}^3}{6W^{3}T}\cdot\left[2+(W^2-1)(1+z)^{3\widetilde{\gamma}}\right]
\end{equation}
The conservation equation for the particle number is:
\begin{equation*}
    (nU^{\mu})_{;\mu}=0
\end{equation*}
which means that $na^{3}=constant$ in the comoving frame.
Therefore the entropy for the whole observable universe in this
model is
\begin{equation}\label{eq:entropy}
    S(z)=(n\sigma)a^{3}=(\frac{a_{0}}{z+1})^{3}S^{0}(z)
\end{equation}
By considering Eqs.(\ref{eq:exp_x}) and (\ref{eq:entropy zero}),
we can obtain

\begin{eqnarray}\label{eq:Ent_z}
    S(z)=&&-\frac{\zeta_{1}T_{1}\theta_{0}^3}{6W^{3}T}\cdot\left(\frac{a_{0}}{1+z}\right)^{3}\cdot\nonumber\\
    &&\left[2+(W^2-1)(1+z)^{3\widetilde{\gamma}}\right]
\end{eqnarray}

With the above expressions, we have established the relation
between the entropy $S$ and the redshift $z$.

Now, we will use 157 gold data points as presented by Riess et
al\cite{AGR} to confront with our model and constrain parameters
$W$ as well as  $\widetilde{\gamma}$. In order to maximize the
following likelihood function (see references\cite{SC,YG}):
\begin{equation}\label{eq:like}
L\propto \exp[-\frac{\chi^{2}}{2}]
\end{equation}
we minimize $\chi^{2}$ which here is expressed as
\begin{equation}\label{eq:kai2}
\chi^{2}=\sum_{i}\left[\frac{\mu_{obs}(z_{i})-\mu_{th}(z_{i})}{\sigma_{i}}\right]^{2}
\end{equation}
where $z_{i}$, $\mu_{th}(z_{i})$, and $\sigma_{i}$ are known from
Gold data. $\mu_{th}(z)$ is defined by
\begin{equation*}
    \mu_{th}(z)=5\lg\left[(1+z)\int_{0}^{z}\frac{dz'}{h(z')}\cdot c\right]+25
\end{equation*}
where $h(z)$ is the reduced Hubble parameter with
$h(z)=H(z)/H_{0}=\theta(z)/\theta_{0}$. Here we have assumed
curvature $k=0$ and $c$ is a constant ( also one of constrained
quantities).

\begin{figure}
  \includegraphics{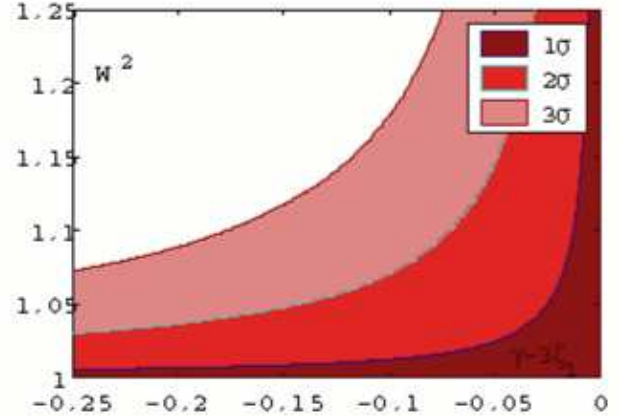}\\
  \caption{The likelihood contours obtained for $W^2$ and $\widetilde{\gamma}$ with 157 gold SNe Ia data analysis.}\label{fig:statefinder11}
\end{figure}

Through the numerical calculations, we find that the best
consistent values can be taken as
\begin{equation*}
    W=-1,\quad \widetilde{\gamma}<0\quad or\quad W\leq-1,\quad \widetilde{\gamma}=0
\end{equation*}
Because the $\widetilde{\gamma}$ has also possessed the constraint
of $\widetilde{\gamma}<0$, we can merely take $W=-1$ then. The
$68.3\%$, $95.4\%$ and $99.7\%$ likelihood contours are shown on
the figure \ref{fig:statefinder11}. The next figure
\ref{fig:statefinder13} is the comparison between experiment data
and the theoretic model estimations,  from which we can see that
the theoretic estimations can well fit the Gold data for the
smaller redshifts.

Taking $W=-1$ into Eq.(\ref{eq:Ent_z}), the expression of entropy
is reduced into
\begin{equation} \label{eq:entropy1}
S(z)=\frac{\zeta_{1}T_{1}\theta_{0}^3}{3T}\cdot\left(\frac{a_{0}}{1+z}\right)^{3}
\end{equation}
in which the entropy density $S^{0}$ does not alternate, but the
entropy $S$ changes as the "volume" $a^3$ varies. Obviously the
entropy of the Eq.(\ref{eq:entropy1}) provides an arrow of time
for cosmic evolution with the meaning
that the entropy of our observable universe is always increasing.

\begin{figure}
  \includegraphics{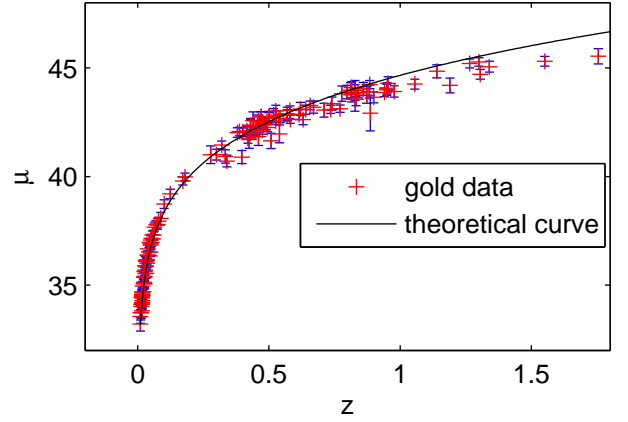}\\
  \caption{The comparison of experiment data and theoretical curve by fitting the 157 gold SNe Ia data}\label{fig:statefinder13}
\end{figure}

For the definition of the parameter $W$,
$W=\widetilde{\gamma}\theta_{0}T_{1}$ , parameter
$\widetilde{\gamma}$ when one takes $W=-1$ becomes
\begin{equation*}
    \widetilde{\gamma}=\gamma-3\zeta_{1}=-\frac{9p_{1}}{\theta_{0}^{2}}=-\frac{p_{1}}{\rho_{0}}
\end{equation*}
Taking the result into Eq.(\ref{eq:rho_W}), we can get
\[\rho=\rho_{0}\] which is a constant. Consequently, pressure
$p$ also takes a constant value. Comparing these results with the
well-known $\Lambda$CDM model, we may conclude that in the viscous
cosmology cases, the best fitting results still favor the
$\Lambda$CDM model. This point can also be reached from the view
point of of statefinder diagnostic pair.

\begin{figure}
  \includegraphics{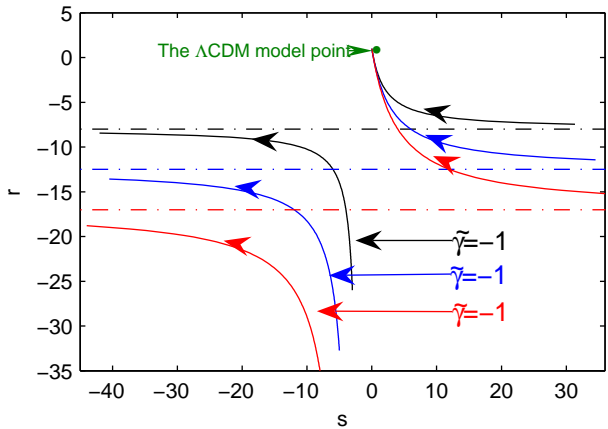}\\
  \caption{On the $s-r$ plane, the trajectory is a hyperbola. The present point is at $r=1$,$s=0$. The arrows indicate the
  cosmic parameters changing tendency, which prefers
 the $\Lambda$CDM model in the later cosmic evolutions.}\label{fig:statefinder14}
\end{figure}

Via using the statefinder diagnostic pair $\{r,s\}$ expressions,
we can obtain directly,
\begin{align}
    q &= \frac{3}{2}V-1  \label{eq:state} \\
    r &= \frac{9}{2}V(\widetilde{\gamma}-1)+1
    \tag{\ref{eq:state}$a$}\\
    s &= \frac{V(\widetilde{\gamma}-1)}{V-1}\tag{\ref{eq:state}$b$}
\end{align}
where $V$ is defined by
\[V=p_{1}(\frac{1}{\rho}-\frac{1}{\rho_{0}})\]
The trajectories is show on the figure \ref{fig:statefinder14}.
When taking the best fitting parameter $W=-1$, we obtain $V=0$,
$r=1$ and $s=0$ that is also the case of the $\Lambda$CDM model.
So it confirms our previous point again that in the viscous cases,
the best fitting model still returns to the conventional
$\Lambda$CDM model.

\section{conclusions and discussions}\label{sec:conclusion}

We mainly discuss the behaviors of some viscous cosmology models
on the statefinder pair plane on the purpose to mimic dark energy
characters, with the hope to demonstrate that cosmic viscosity can also
play the role as
a possible candidate for dark energy. To this aim, we are first to
give out the formulas for viscous cosmology statefinder pair
expressions (\ref{eq:Q}). After introducing the relative changing
rate $\varsigma$, the global evolution of density is found to
relax as more stable in the viscous model situations. At the same
time, the trajectories of viscous universe statefinder pair on
$s-r$ plane become quite different from non-viscous cases as
table \ref{tab:table2} 
\begin{table}
\caption{\label{tab:table2} Relationship between models and figures
}
\begin{ruledtabular}
\begin{tabular}{lcr}
the $s-r$ plane&$\zeta=\zeta_{0}$&$\zeta=\zeta_{1}\dot{a}/a$\\
\hline
$\Lambda$CDM &Fig.\ref{fig:statefinder2} & $\otimes$\footnotemark[1]\\
Quiessence & Fig.\ref{fig:statefinder4} & $\otimes$\footnotemark[1]\\
Variable DE EoS & $\otimes$\footnotemark[1] & Fig.\ref{fig:statefinder14}\\
\end{tabular}
\end{ruledtabular}
\footnotemark[1]{$\otimes$ denotes no figures}
\end{table}

Particularly for the Quiessence model, the singular property of
the `phantom divide' $w_{0}=-1$ can be clearly demonstrated on the
statefinder diagnostic pair plane by the completely different
trajectories to discriminate themselves. And the directions of the
evolution of the trajectories on the statefinder diagnostic pair
planes for the three models all point to the point of $r=1$,$s=0$
($\Lambda$CDM model preferred), that is the universe favors simple
in the later evolution stage with scale largely expanded. The
$\Lambda$CDM cosmology is simply consistent with the current
astrophysics observations, especially the cosmic late time
accelerating expansion,  but the cosmological constant has been
puzzling ever since.

 Additionally, we
also have a try to describe the entropy of the viscous cosmology
system. After adopting the EoS (\ref{eq:pq}) and the bulk viscosity
$\zeta=\zeta_{1}H$, we deduce out the concrete expression for the
entropy. Then the 157 gold data from the supernova observation are
used to constrain parameters $W$ and $\widetilde{\gamma}$, and we
therefore get the most favorable parameter: $W=-1$. Further we find
that the entropy of the universe is always increasing with cosmic
evolution, which is consistent with the thermodynamics arrow of
time.

Observational cosmology across this century  has challenged our
naive physics models, and with the anticipated advent of more
precious data we have the chance to understand or uncover the
universe mysteries by more practical modelling. Quite possibly we
will get more hints to unveil the cloudy cosmological constant
puzzle. In the simple constant bulk viscosity case (a proto type
or a toy cosmic media model) as demonstrated in section three the
vacuum energy density can be shifted by the bulk media viscosity
to arrive at an effective vacuum energy density (EVED) or we may
say that the constant viscosity can tune the cosmological constant
in a sense if we have possessed a suitable cosmic media model. We
expect more encouraging work on non-perfect fluid cosmic concord
models to come soon and we believe this line of trying can
contribute us new understandings to the mysterious dark side of
our complicated but observable and conceivable universe.

\section*{ACKNOWLEDGEMENTS}
We thank Prof. S.D. Odintsov for the helpful comments with reading
the manuscript, and  Profs. I. Brevik and L. Ryder for lots of
discussions. This work is supported partly by NSF and Doctoral
Foundation of China.

\end{document}